\documentclass[structabstract]{aa}  
\usepackage{graphicx}
\usepackage{epsfig}
\usepackage{lscape}
\pdfoutput=1
\usepackage{txfonts}
\begin{document}
%
   \title{Suzaku and SWIFT-BAT observations of a newly discovered 
   Compton-thick AGN}


   \author{P. Severgnini,
          \inst{1}
	  A. Caccianiga,
          \inst{1}
	  R. Della Ceca,
          \inst{1}
	  V. Braito,
	  \inst{2}
	  C. Vignali,
          \inst{3,}
          \inst{4}
	  V. La Parola
	  \inst{5}
          \and
	  A. Moretti
	  \inst{1}
          }

   \institute{INAF-- Osservatorio Astronomico di Brera, via Brera 28, 20121 Milano, Italy\\
              \email{paola.severgnini@brera.inaf.it}
	      \and
               Department of Physics and Astronomy, Leicester University, 
               Leicester LE1 7RH, UK
	       \and    
               Dipartimento di Astronomia, Universita' degli Studi di Bologna, 
	       via Ranzani 1, 40127 Bologna, Italy
	       \and 
	       INAF--Osservatorio Astronomico di Bologna, via Ranzani 1, 
	       40127 Bologna, Italy
	       \and
	       INAF--Istituto di Astrofisica Spaziale e Fisica Cosmica di Palermo, 
	       via U. La Malfa 153, 90146 Palermo, Italy}

   \date{Received ...; accepted ...}

 
  \abstract
     {Obscured AGN are fundamental to understand the history of Super Massive Black Hole growth
and their influence on galaxy formation. However, the Compton-thick  AGN 
($N_\mathrm{H}$$>$10$^{24}$
cm$^{-2}$) population is basically unconstrained, with less than few dozen confirmed
Compton-thick AGN found and studied so far. A way to select heavily obscured AGN is to
compare the X-ray emission below 10 keV (which is strongly depressed) 
with the emission from other bands less affected by the absorption, i.e.
the IR band. To this end, we have cross-correlated the 2XMM catalogue with the IRAS 
Point Source catalogue and, by using the X-ray to infrared flux ratio and 
X-ray colors, we selected a well defined sample of Compton-thick AGN 
candidates at z$<$0.1.}
      {The aim of this work is to confirm the nature and to study one of these local
Compton--thick AGN candidates, the nearby (z=0.029) Seyfert 2 galaxy IRAS
04507+0358, by constraining  the amount of intrinsic absorption ($N_\mathrm{H}$) and
thus the intrinsic luminosity.} 
   {To this end we obtained deep (100 ks) Suzaku observations (AO4
call) and performed a joint fit with SWIFT--BAT data. We analyzed 
{\it XMM-Newton},
Suzaku and SWIFT--BAT data and we present here the results of this 
broad--band (0.4-100 keV)
spectral analysis.}
    {We found that the broad--band X--ray emission of IRAS 04507+0358 requires 
   a large amount of absorption (larger than 10$^{24}$ cm$^{-2}$) to be well reproduced, 
   thus confirming the Compton--thick nature of this source. In particular,  the most probable
   scenario is that of a
   mildly ($N_\mathrm{H}$$\sim$1.3-1.5$\times$10$^{24}$ cm$^{-2}$, L(2--10 keV)$\sim$5-7$\times$10$^{43}$ erg s$^{-1}$) 
   Compton--thick AGN.
 }
   {}


        \authorrunning{Severgnini et al.}
              
   \maketitle
%

\section{Introduction}

On the basis of the most accredited X-ray background synthesis models (Gilli et al.
\cite{gilli07}, Treister et al. \cite{treister}, Ballantyne et al
\cite{ballantyne06}), obscured AGN ($N_\mathrm{H}$$>$10$^{22}$ cm$^{-2}$) dominate the
entire AGN population. For this reason, their space density at  different
redshifts is one of the main ingredients in setting the evolutionary properties
of Super Massive Black Hole (SMBH). Unfortunately, the absorption of the
obscuring medium (composed of gas and dust) along the line of sight  does not
allow us to easily detect them and study their nuclear properties in the
UV/optical bands and in the soft X-ray bands (energy below few keV).  This is
particularly true for the most  obscured sources, the  so called
Compton--thick AGN, that are hidden by a large amount of  circum-nuclear
obscuring matter along the line of sight (column density of $N_\mathrm{H}$$>$10$^{24}$
cm$^{-2}$).

Compton--thick AGN are generally divided in two main classes of sources: mildly
Compton-thick with $N_\mathrm{H}$ of the order of a few times 10$^{24}$ cm$^{-2}$ and
heavily Compton-thick with $N_\mathrm{H}$ above $\sim$10$^{25}$ cm$^{-2}$.  The primary
radiation is strongly suppressed at low energies in case of mildly Compton-thick
AGN, emerging only above 10 keV, while in the case of heavily Compton-thick AGN
the primary radiation is strongly depressed also above 10 keV, due to the
compton down-scattering effect (Matt \cite{matt96}).   In both classes of
sources, the spectrum below 10 keV is dominated by a pure reflection component
(i.e. the continuum emission  reflected by the putative torus), that is much
fainter than the direct one.  For this reason, even for intrinsically luminous
objects, it is generally difficult to accumulate enough counts below 10 keV to
allow us a reliable spectral analysis and thus to assess the  nature of these
sources.  Furthermore, even  when the spectral analysis is possible, the
absorption cut-off is only marginally detectable below 10 keV (or completely
outside the observed energy window). This fact does not allow to  correctly 
measure the intrinsic absorption and thus to estimate the nuclear luminosity.
Moreover, as we will show in this paper, in spite of the different values of
intrinsic $N_\mathrm{H}$, the shape of Compton-thin AGN with  5$\times$10$^{23}$
cm$^{-2}$$<$$N_\mathrm{H}$$<$10$^{24}$ cm$^{-2}$ and Compton-thick AGN spectra below 10
keV may appear very similar  and indistinguishable, specially in cases of low
counting statistics. This makes even more difficult to adequately probe the
heavily obscured AGN population. Often the presence of Compton thick matter is
inferred through indirect arguments, such as the presence of a strong iron
emission line at 6.4 keV; for high values of $N_\mathrm{H}$, the equivalent width of this
line is expected to be high, reaching values as a few keV (Matt et al.
\cite{matt96}; Murphy \& Yaqoob \cite{murphy}). 

To properly set the presence of a Compton-thick source  and measure the amount
of absorption, hard X--ray data above 10 keV are needed (Comastri et al.
\cite{com10}). These data allow to detect the absorption cut-off in the very
hard X--ray domain typical of large amount of intrinsic $N_\mathrm{H}$
($>$10$^{24}$ cm$^{-2}$). For the vast  majority of Compton--thick AGN without
high energy data, only a conservative lower limit on the intrinsic column
density and nuclear luminosity can be placed. Extreme examples of such
limitations are NGC 6240 (Vignati et al. \cite{vignati}), Mrk 231 (Braito et al.
\cite{braito04}) and Arp~299 (Della Ceca et al. \cite{rdc02}), where only
Beppo-SAX PDS (energy range 15-300 keV) observations were able to reveal the
intrinsic AGN emission, allowing the direct measure of the $N_\mathrm{H}$ and
the intrinsic AGN luminosity. 

A reliable estimate of the number of Compton--thick AGN requires first of all
the use of efficient methods able to select large and possibly complete samples
of Compton--thick AGN candidates. As a second step, broad-band X--ray  spectra
(from a few to hundreds keV) are needed to confirm their Compton--thick nature.
However, we lack such a statistical sample and, more importantly, only a limited
fraction of  the putative Compton thick AGN have been confirmed through
broad--band X--ray spectroscopy (see Comastri et al. \cite{com04}; Della Ceca et
al.\cite{rdc08}, Awaki et al. \cite{awaki09}, Teng et al. \cite{teng}, Braito et
al. \cite{braito09}). For this reason, indirect arguments are generally used to
estimate their density in the local Universe  (Della Ceca et al.
\cite{rdc2008}). 

In Severgnini et al. (\cite{sev}) we found that the (2--10 keV) to 24$\mu$m
flux ratio  vs. X--ray colors can be used as an efficient technique to select
Compton--thick candidates, at least in the local Universe. In particular, after
having cross--correlated  the IRAS Point Source Catalog v2.1 (PSC)  with the
bright end (F$_{2-10 keV}$$>$10$^{-13}$ erg cm$^{-2}$ s$^{-1}$) of the
incremental version of the 2XMM  catalogue (Watson et al. 2009), we found that
85\% of the 46 extragalactic sources with F(2-10 keV)/($\nu_{24 \mu
m}$F24$\mu$m)$<$0.03  and  HR4\footnote{HR4 is defined using the two
following bands: 2-4.5 keV and 4.5--12 keV: HR4=$\frac{CTS(4.5-12 keV) -
CTS(2-4.5 keV)}{CTS(4.5-12 keV) + CTS(2-4.5 keV)}$, where CTS are the vignetting
corrected count rates in the energy ranges reported in bracket. See Watson et
al. \cite{wat09} for details.}$>-$0.1 have X--ray
properties resembling those of a typical Compton-thick AGN. Ten of these are
newly discovered Compton-thick candidates. In order to confirm the nature of
these objects, we have both cross--correlated our Compton--thick AGN candidates
with the SWIFT--BAT catalogue (54 months of observations,
Cusumano et al. \cite{cusumano10})  and obtained Suzaku observations for
two of them (IRAS 04507+0358 and MCG-03-58-007).  Here we present the results
obtained with the broad--band X--ray analysis (Suzaku plus SWIFT--BAT) of IRAS
04507+0358, the first target observed by Suzaku, while the results obtained
with the cross-correlation with the SWIFT--BAT catalogue will be presented in a
forthcoming paper (Severgnini et al. in prep.). Throughout this paper we use
the cosmological parameters H$_0$=71 km s$^{-1}$ Mpc$^{-1}$, 
$\Omega_{\lambda}$=0.7 and $\Omega_{M}$=0.3.

\begin{figure*}
\begin{center}
\psfig{figure=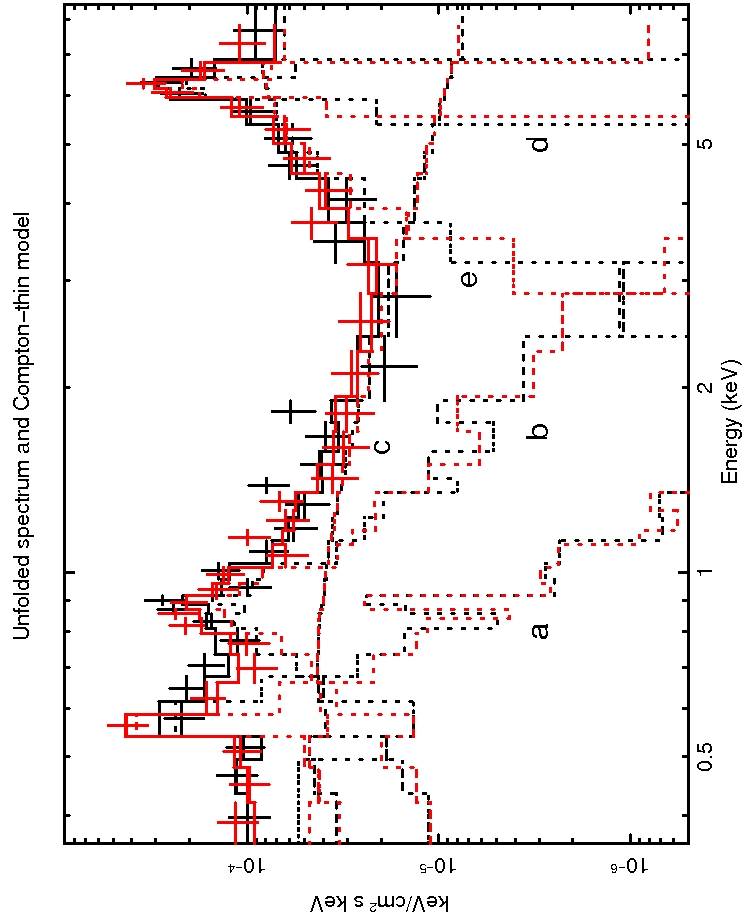,height=9.0cm,width=7.0cm,angle=-90}
\psfig{figure=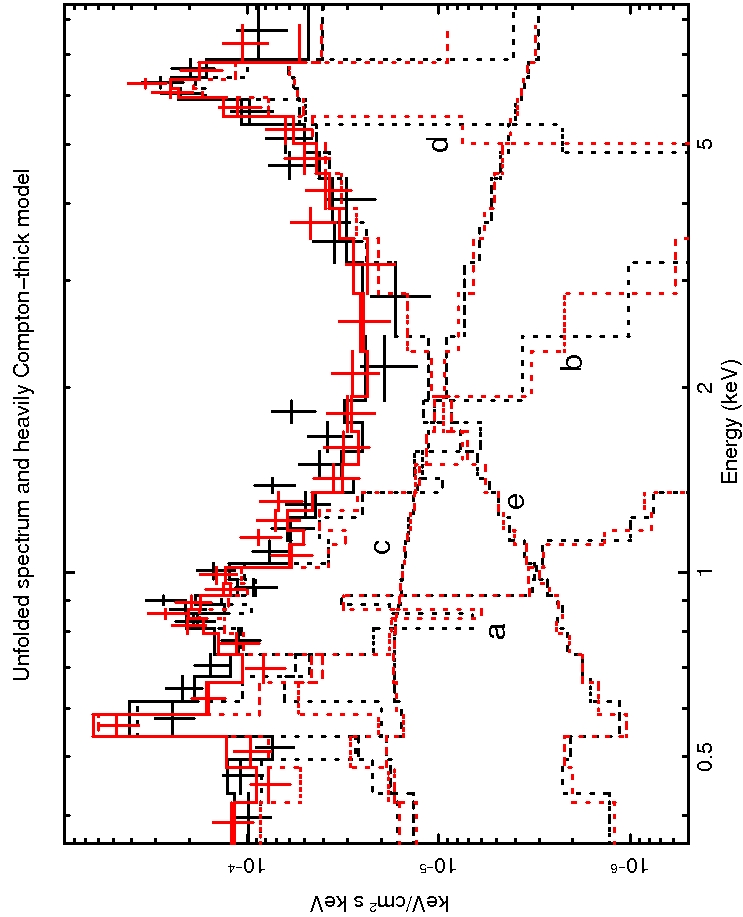,height=9.0cm,width=7.0cm,angle=-90}
\psfig{figure=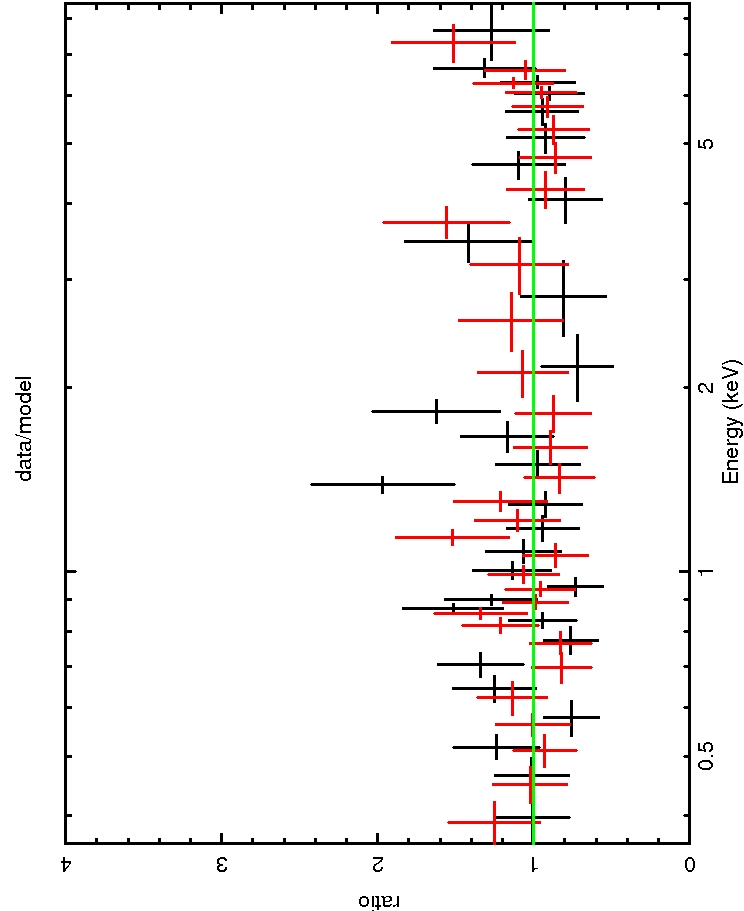,height=9.0cm,width=6.0cm,angle=-90}
\psfig{figure=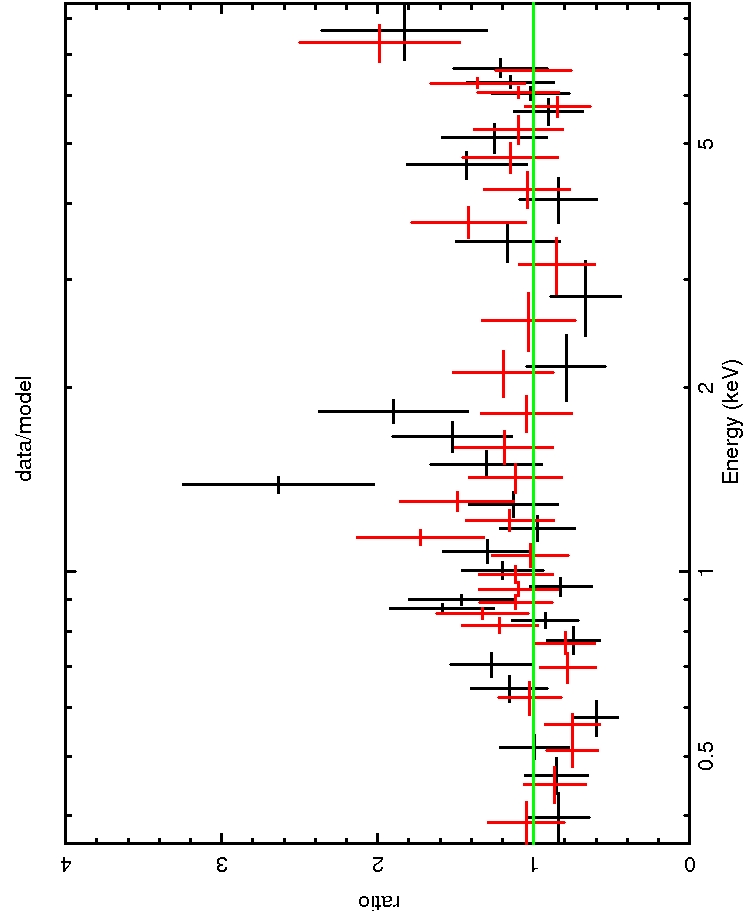,height=9.0cm,width=6.0cm,angle=-90}
\label{Fig 1}
\caption{{\it XMM--Newton} (MOS1 -- black data points in the electronic
version  and MOS2 -- red data points in the electronic version) unfolded
spectrum of IRAS 04507+0358  fitted with a Compton-thin (left panel) and a
Compton-thick model (right panel). In the lower panels  the ratio between data
and best--fit models are shown. In the Compton-thin hypothesis, the data are
best reproduced by two thermal emission components (a, b) plus  a
power--law absorbed only by the Galactic column density  (scattering component,
c) and a power--law  absorbed by a high column density 
($N_\mathrm{H}$=(4$\pm{2}$)$\times$10$^{23}$ cm$^{-2}$) absorber (e).  The two
power-laws have the same $\Gamma$ fixed to 1.9.  A prominent iron line is also
present (d).
A similarly good fit  could be obtained by using a cold reflection component
instead of the absorbed power--law (e component in the left panel).}
\end{center}
\end{figure*}

\section{IRAS 04507+0358}

This source is optically classified as Seyfert~2 galaxy at z=0.029 (Strauss et
al. \cite{strauss}). No clear signs of starburst activity have been detected 
(Cid Fernandes et al. \cite{cid}). From the  IRAS--PSC catalogue fluxes, we
calculated the infrared (8--1000 $\mu$m) luminosity following the prescription
reported in Sanders \& Mirabel (\cite{sanders}) and Risaliti et al.
(\cite{risaliti}). We found L$_{IR}$$\simeq$2.7$\times$10$^{44}$ erg
s$^{-1}$$\simeq$7$\times$10$^{10}$ L$_{\odot}$, in agreement with the estimate reported
in Fraquelli et al. (\cite{fraquelli}).

\subsection{{\it XMM-Newton} archival data}  The 2--10 keV spectrum used in the
analysis described here  (Observation ID = 0307000401, $\sim$11 ksec of good
time exposure in the MOS camera only) was taken from the {\it XMM--Newton}
archive as one of the products of the 2XMM catalogue (Watson et al.
\cite{wat09}). The data are grouped with a minimum of 20 counts per  channel. 
The X--ray emission of IRAS 04507+0358  (F(2--10 keV)$\sim$10$^{-12}$ erg
cm$^{-2}$ s$^{-1}$) could be explained with two formally acceptable models: a
transmission dominated ($N_\mathrm{H}$$<$10$^{24}$ cm$^{-2}$) and a reflection
($N_\mathrm{H}$$>>$10$^{24}$ cm$^{-2}$) dominated scenario (see Fig. 1). In both
cases, two thermal components are required to fit the softer part of the 
spectrum (below 2 keV). We used two {\it mekal} models (Mewe et al.
\cite{mewe85}, \cite{mewe86};  Liedahl et al. \cite{lie}) with the abundances of
Wilms et al. (\cite{wilms}),  typical of the inter--stellar medium. We found 
{\it k}T$_1$=0.15$^{+0.04}_{-0.05}$ and {\it k}T$_2$=0.72$^{+0.09}_{-0.06}$ keV
(if we use the Compton--thin model) and  {\it k}T$_2$=0.66$^{+0.03}_{-0.02}$ keV
(if we use the Compton--thick model).   These thermal components are most
probably associated to star-forming  activity. As discussed in Section 3, we
have estimated a star formation rate (SFR) of  about 10 M$_\odot$/yr in our
source on the basis of the infrared luminosity. For these values of SFR, the
presence of a multi--temperature components in the X--ray spectra is generally
observed (e.g. M~82, Persic et al. \cite{persic}). 

As for the data above 2 keV, a clear roll-over is present in the spectrum, 
which, in the Compton-thin hypothesis, could be interpreted as a
signature of the photoelectric cutoff. We used two power--laws with the same
photon index fixed to $\Gamma$=1.9 (Reeves \& Turner \cite{reeves00};  Caccianiga
et al. \cite{caccia04}; Page et al. \cite{page}) to model the scattered and the
primary  X--ray emission. While the scattering component was absorbed only
through  the Galactic  column density, we estimated an intrinsic column density
that absorbs the primary X--ray emission of $N_\mathrm{H}$=(4$\pm{2}$)$\times$10$^{23}$
cm$^{-2}$ ($\chi^{2}$/d.o.f.=41.8/52, left panel of Fig. 1). A similarly good
fit could be obtained also by replacing the absorbed power--law with a
reflection component ({\it pexrav} model, Magdziarz \& Zdziarski
\cite{mag}). We call this model the "heavily Compton-thick hypothesis"
($\chi^{2}$/d.o.f.=53.2/53, right panel of Fig. 1).  Even in this case,  we used
the same slope ($\Gamma$=1.9) for both the scattered and the reflected 
component. We fixed the reflection fraction (defined by the subtending solid 
angle of the reflector R=$\Omega$/2$\pi$) equal to 1 and the inclination angle
to the mean value of 60$^{\circ}$. We kept the abundances of  Wilms et al.
(\cite{wilms}). Independently from the assumed model for the underlying
continuum, a strong  narrow Fe line is detected. We fixed the energy line at 6.4
keV  and we found an EW=1.1$^{+1.4}_{-0.4}$ keV ($\sigma$=0.2$\pm$0.1 keV)  in
the Compton--thin model and EW=2.2$^{+0.8}_{-0.6}$ keV
($\sigma$=0.3$^{+0.4}_{-0.1}$ keV) in the heavily Compton-thick hypothesis. 

Apart from  the strong iron line that could be considered as an indirect hint
for the presence of a Compton--thick source, this analysis shows that the
{\it XMM-Newton} data alone can not distinguish between a Compton-thin or
Compton-thick scenarios. With the aim of assessing the nature of this object and
estimating its intrinsic properties (e.g. $N_\mathrm{H}$ and luminosity), we have 
obtained 100 ks of Suzaku observation in the A04 call.   We have combined here
this Suzaku data with SWIFT--BAT spectrum accumulated during the first 
54--months of observations  (Cusumano et. \cite{cusumano10}).

\subsection{Suzaku data}

IRAS 04507+0358 was observed by the Japanese X-ray satellite Suzaku (Mitsuda et
al. \cite{mitsuda}) for 100 ks at the beginning of September 2009.  At the time
of these observations only three of the four X-ray Imaging Spectrometers (XIS,
Koyama et al. \cite{koyama}) were working: two front-illuminated (XIS0 and XIS3)
and one back-illuminated (XIS1) CCDs. For our analysis we used the 0.4--8 keV
and 0.4--10 keV data obtained by XIS1 and  XIS0+XIS3 respectively, combined with
the HXD--PIN data. This latter is a non imaging hard X-ray detector (Takahashi
et al. \cite{taka}) covering the 12--70 keV energy band. IRAS 04507+0358 was
placed at the  HXD nominal pointing.  Cleaned event files were filtered
with the standard screening\footnote{The screening procedure filter all events
within the South Atlantic Anomaly (SAA) as well as with an Earth elevation angle
(ELV) $<$5$^{\circ}$ and Earth day-time elevation angles (DYE\_ELV) less than
20$^{\circ}$. Furthermore also data within 256s of the SAA were excluded from
the XIS and within 500s of the SAA for the HXD. Cut-off rigidity (COR) criteria
of $>$8GV for the HXD data and $>$6GV for the XIS were used.}. The effective
exposure time after data cleaning are 83.7 ks for each of the XIS source's 
and 77.6 ks for the  HXD-PIN.

{\bf XIS spectra:} were extracted from a circular region of 2.3 arcmin of
radius  centered on the source. Background spectra were extracted from two
circular regions with the same radius of the source region but offset from the
source and the calibration sources. The XIS response (rmfs) and  ancillary response (arfs) files were
produced using the latest calibration files
available and the ftools tasks {\it xisrmfgen} and  {\it xissimarfgen},
respectively. The net count rates observed with the three XIS
in the 0.4-10 keV  band are 0.034$\pm{0.001}$ cts/s (XIS0), 0.048$\pm{0.013}$
cts/s (XIS1), 0.040$\pm{0.001}$ cts/s (XIS3).

The spectra from the two front-illuminated CCD  were then
combined, while the back--illuminated CCD spectrum was kept separate and
then fitted simultaneously. The net XIS source spectra were then binned in order
to have a minimum signal--to--noise ratio (S/N) $\simeq$4 in each energy bin. 

{\bf HXD--PIN spectrum:} At the time of writing, two instrumental background
(called non--X--ray background, NXB) files have been released, the {\it quick}
and the {\it tuned} one. We tested  both of them and we found that the {\it tuned}
background count rate in the 15--70 keV is 5\% higher  than the count rate of the
quick one in the same energy range. To estimate which is the event file that 
provides
the most reliable estimate of the real NXB, we compared them with the data taken
during periods of Earth occultation in the same energy range, 15-70 keV. Since
the Earth is known to be dark in hard X--rays and during the Earth occultation
we do not measure the emission of the source and of the cosmic X--ray
background, the occulted data should give a good representation of the actual
NXB rate. Thus, after having corrected the Earth occultation data for the
dead-time and taken into account only the events with an Earth elevation angle
lower than -5$^{\circ}$, we extracted the Earth occultation spectrum and
light--curve in the 15--70 keV an 15--50 keV, respectively.

The comparison between the occulted data and the {\it quick} (bkgA) and {\it
tuned} (bkgD) backgrounds is shown  in Fig. 2. The {\it tuned} background  level
is systematically higher than the {\it quick} background level and than the
Earth occultation level. This difference becomes larger at E$>$35 keV. The
light-curve comparison is shown in Fig. 3.  The {\it tuned} background spectrum
and light-curve are nearly 7\% above the data taken during the Earth
occultation.

As a final check, we compared Suzaku flux with Swift-BAT data. We found that
with the {\it quick}  background, the Suzaku flux is close to the Swift one, 
i.e. F(15--70 keV)$_{Suzaku}$=1.9$\times$10$^{-11}$ erg cm$^{-2}$ s$^{-1}$ and
F(15--70 keV)$_{BAT}$=1.8$\times$10$^{-11}$ erg cm$^{-2}$ s$^{-1}$ (assuming the
same model for the two instruments). We thus
decided to use the {\it quick} background file and combined it with  the  cosmic
X--ray background, the latter was parametrized by us using the prescription of
Boldt (\cite{boldt}) and Gruber et al. (\cite{gruber}). The background-corrected
count rate in the 15$-$70 keV is 0.033 cts/s ($\sim$8.3\% above the background
level). For the spectral analysis, we rebinned the HXD-PIN spectrum in order to have a signal-to-noise ratio of 4 in each energy bin.

\subsection{SWIFT--BAT data}

The SWIFT--BAT data were processed using the Bat\_Imager software (Segreto et
al. \cite{seg}). IRAS~04507+0358 was detected with a S/N$\sim$12 in the 14-150
keV energy range. In order to produce the spectrum of IRAS~04507+0358 integrated
over the 54 months of survey data, we extracted the rates and the relevant
errors from the pixel corresponding to the best BAT position in the all-sky maps
produced in eight energy bands (14-20 keV, 20-24 keV, 24-35 keV, 35-45 keV,
45-60 keV, 60-75 keV, 75-100 keV, 100-150 keV). The spectrum was analyzed using
the BAT spectral redistribution matrix distributed with the BAT CALDB. 
Further details can be found in Cusumano et al. (\cite{cusumano09}).

\begin{figure}[]
\begin{center}
\psfig{figure=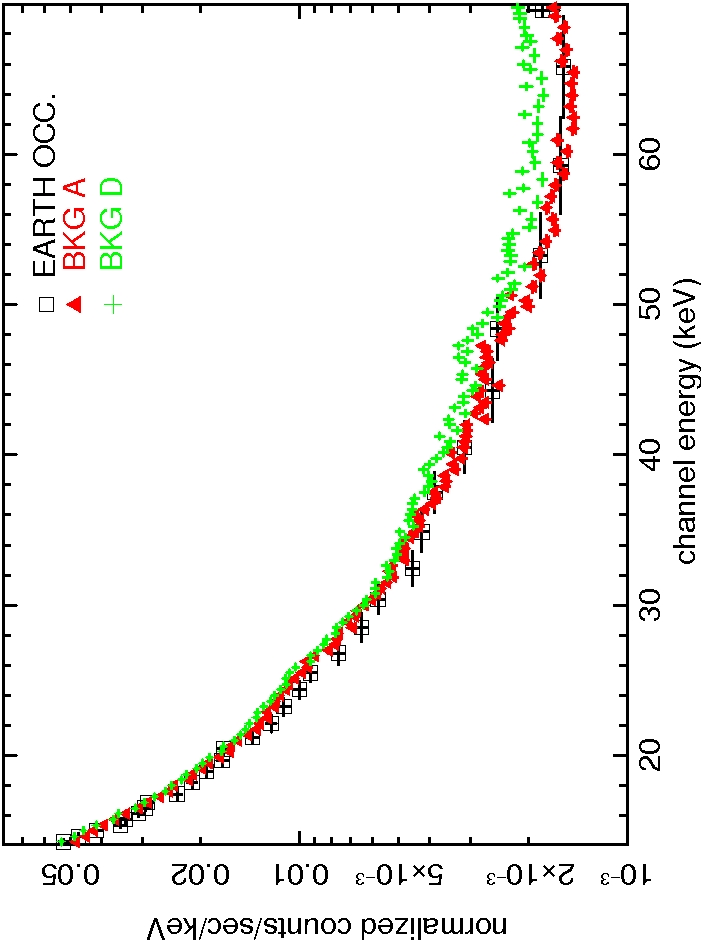,height=9.0cm,width=7.0cm,angle=-90}
\label{Fig. 2}
\caption{Comparison between the {\it quick} (bkgA, red triangles) and 
the {\it tuned} (bkgD, green crosses) backgrounds with the  Earth occultation spectra (black squares). 
The BkgD spectrum is above both the bkgA
and the Earth data. The {\it tuned} background clearly overpredicts the real NXB, in particular
above 35 keV.}   
\end{center} 
\end{figure}

\begin{figure}[]
\begin{center}
\psfig{figure=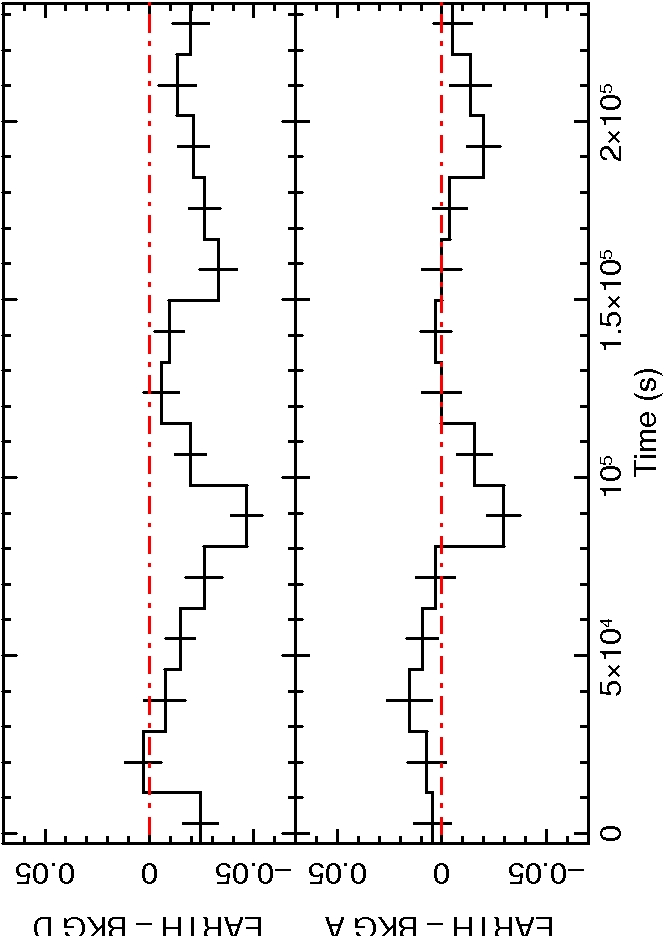,height=9.0cm,width=7.0cm,angle=-90}
\label{Fig. 3}
\caption{Comparison between light curves extracted in 
the 15-50 keV energy range. The Earth light curves are corrected for 
the dead time. In the upper panel we show the difference between the 
Earth and the {\it tuned} background (bkgD) light curves, while in the
lower panel we show the difference between the Earth and the {\it quick}
background (bkgA) light curves.} 
\end{center} 
\end{figure}

\begin{figure} \begin{center} \label{Fig. 4}
\psfig{figure=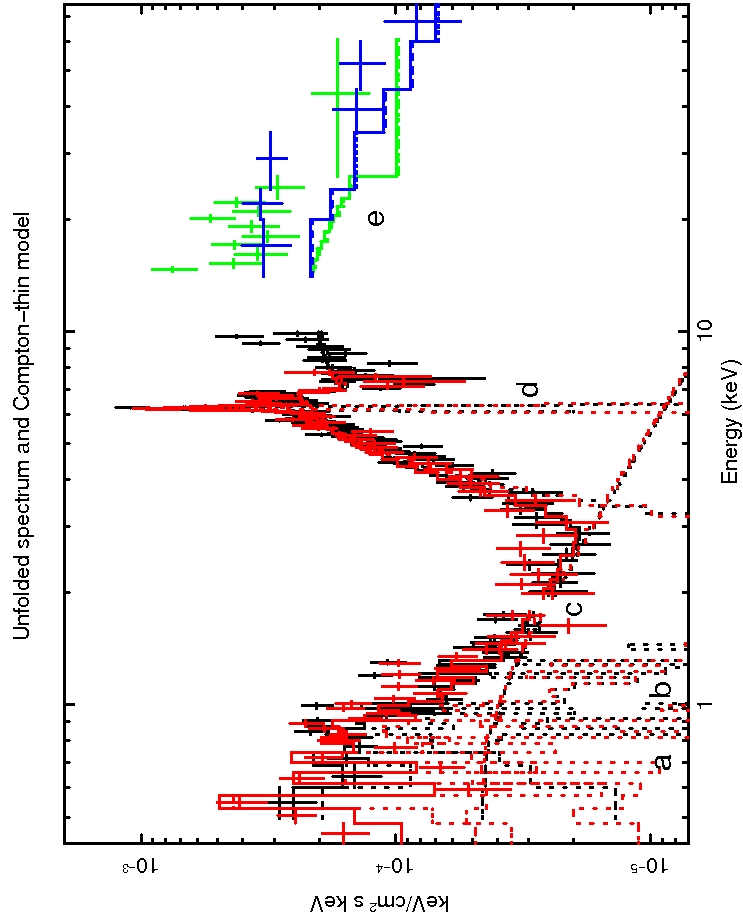,height=9cm,width=7cm,angle=-90}
\psfig{figure=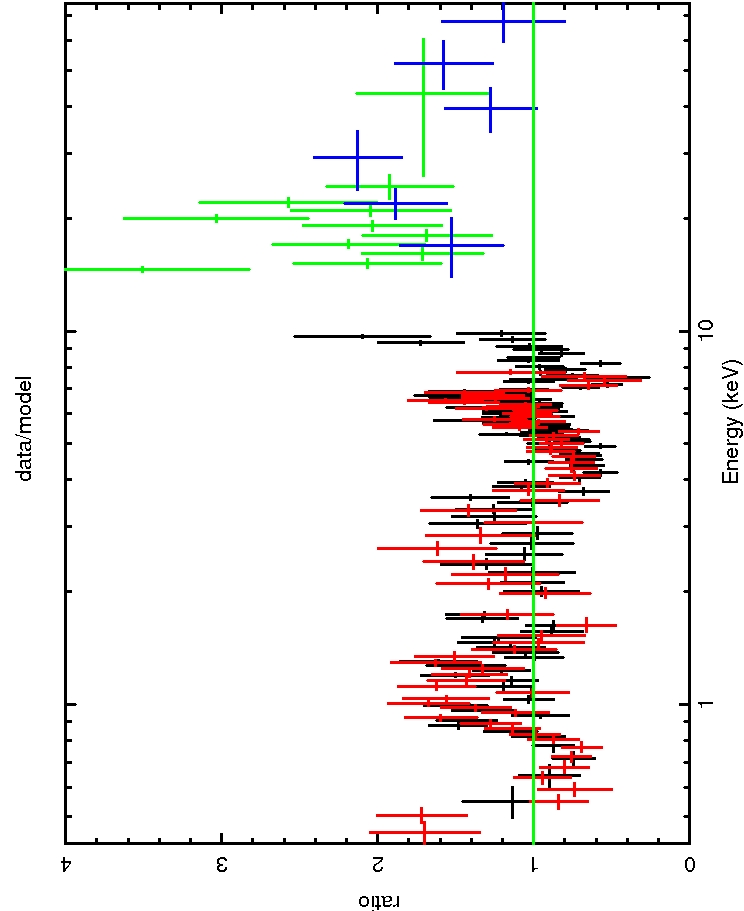,height=9cm,width=6cm,angle=-90}
\caption{Suzaku XIS-FI (black data points in the electronic version), 
XIS-BI (red in the electronic version), HXD--PIN  (green in the
electronic version) and SWIFT--BAT  (blue in the electronic version) data
of IRAS~04507+0358.  {\it Upper panel:} Unfolded Compton--thin model. The model
consists of: two thermal emission components ({\it a} and {\it b} components), a
scattered component ({\it c}),
a Fe emission line at $\sim$6.4 keV ({\it d}) and an absorbed power--law 
 with a $N_\mathrm{H}$ fixed to 4$\times$10$^{23}$ cm$^{-2}$ ({\it e}). 
{\it Lower panel:} Ratio between data
and the Compton--thin best--fit model.}   
\end{center}   
\end{figure}

\subsection{Broad band X--ray spectral fitting} 

\begin{figure}[th]
\begin{center}
\psfig{figure=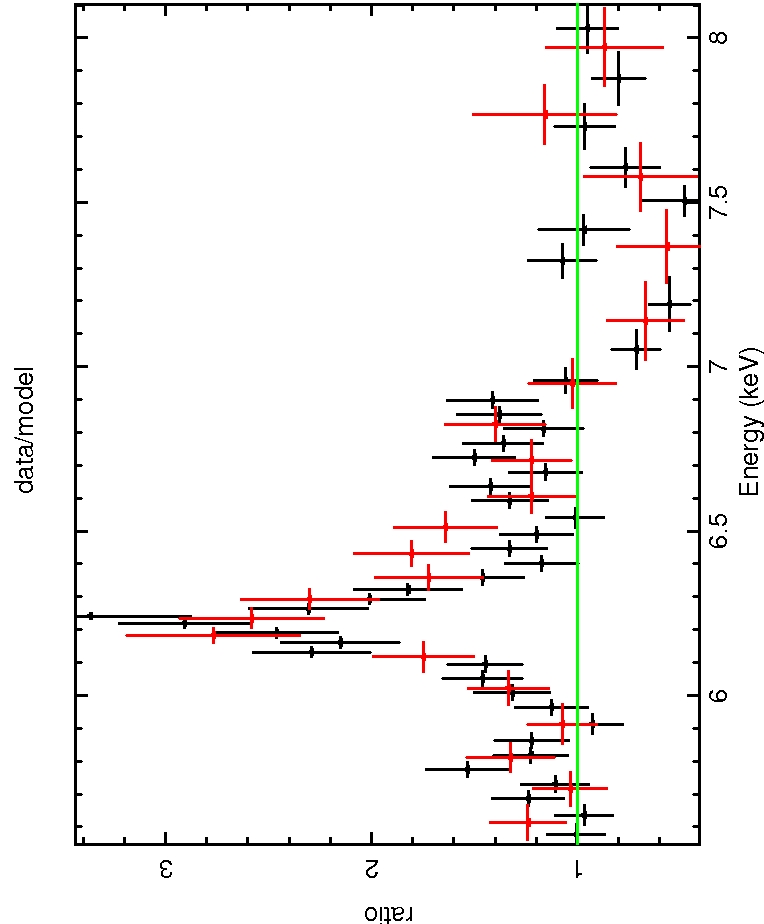,height=9cm,width=7cm,angle=-90}
\label{Fig. 5}
\caption{Residuals, with respect to a single power-law, of the XIS data/model 
 (XIS-FI - black data points  in the electronic version and  
XIS-BI-red points in the electronic version) of IRAS~04507+0358 
at energies close to the Fe band (observed frame). 
No iron line is included in the model.}
\end{center}  
\end{figure}

The XIS, the HXD--PIN and the BAT spectra have been fitted  simultaneously covering a wide
energy range (0.4--100 keV) using Xspec version 12.5.0. 
We tied together XIS, HXD and BAT parameters. We left free to vary the BAT/XIS 
cross-normalization, while for the HXD/XIS instruments we assumed a
cross--calibration of 1.18 (Manabu et al. 2007; 
Maeda et al. 2008\footnote{http://www.astro.isas.jaxa.jp/suzaku/doc/suzakumemo/suzakumemo-2007-11.pdf; \\
http://www.astro.isas.jaxa.jp/suzaku/doc/suzakumemo/suzakumemo-2008-06.pdf}). 
In all the models described here we have used the abundance of Wilms et al. 
(\cite{wilms}) and the 
Galactic hydrogen column density along the line of sight (from Dickey \& Lockman
\cite{dickey}), $N_\mathrm{H}$(Galactic)=6.7$\times$10$^{20}$ cm$^{-2}$.

Since the aim of this work is to explore the physical properties of the nuclear 
regions of IRAS~04507+0358 (e.g. the $N_\mathrm{H}$ of the obscuring matter and the
de-absorbed X--ray luminosity of the AGN), a detailed analysis of the spectrum
below 2 keV  will be not discussed here. Starting from the results obtained with
{\it XMM--Newton} data, we have fitted this part using two thermal components.  We
left the temperatures free to vary and we found that the two temperatures are
in  good agreement with those found with {\it XMM-Newton} data: {\it
k}T$_1$=0.14$^{+0.05}_{-0.01}$ and {\it k}T$_2$=0.73$^{+0.07}_{-0.04}$. 

As for the {\it XMM-Newton} spectra, to fit the data  above 2 keV,  we used
a simple basic model for obscured AGN, i.e. two power--laws with the same slope.
One of the two power--laws is absorbed only by Galactic column density and
represents the scattered component; the second power-law represents the primary
X--ray emission seen through the cold absorber (the putative torus).  To
reproduce this absorbed power--law, we used the model by Yaqoob \cite{yaqoob97}
({\it plcabs} in Xspec), which partially takes into account for the Compton
down-scattering.  The Fe emission line was modeled with a Gaussian profile.\\ 

{\bf Compton-thin hypothesis:} As a first step, we tested the Compton--thin 
hypothesis by fixing the slope of the power--laws to $\Gamma$=1.9 and the column
density to  $N_\mathrm{H}$=4$\times$10$^{23}$ cm$^{-2}$, i.e. the values found
with the {\it XMM--Newton} data (see Fig. 4, $\chi^{2}$/d.o.f.=527.5/194). This
model provides an overall good representation of the XIS data,  but it
under-predicts the HXD-PIN and the BAT data. We left the photon index and the
column density free to vary and we found $\Gamma$=1.8$^{+0.3}_{-0.2}$ and 
$N_\mathrm{H}$=5.5$^{+0.2}_{-0.8}$$\times$10$^{23}$ cm$^{-2}$
($\chi^{2}$/d.o.f.=334.95/193); however also  this model does not provide a
good  representation of the emission above 10 keV. Our first conclusion is that
the Compton-thin hypothesis is clearly ruled out by the HXD--PIN and BAT
data.\\

{\bf Mildly Compton-thick hypothesis:}  To account for the prominent Fe emission
line in the spectrum and for the excess detected above 10 keV, we added to
the model a component, which represents the emission  reflected from neutral
material into our line of sight. In particular, we added the {\it pexrav} model
leaving open the possibility that the reflected emission could be partially
obscured by the torus itself. We fixed the inclination angle to its default
value (i = 60$^{\circ}$) and the normalization equal to that of the intrinsic powerlaw. 
With the addition of this new component, that we found to be absorbed
($N_\mathrm{H}$=(3.3$\pm{0.2}$)$\times$10$^{23}$ cm$^{-2}$), the model provides a
relatively good
fit to the 0.4--100 keV spectrum ($\chi^{2}$/d.o.f. = 265.5/190). We found a
BAT/XIS cross-normalization of 1.1$^{+0.1}_{-0.2}$.  The primary X--ray
continuum has an intrinsic slope of $\Gamma$=2.4$\pm$0.5 and is absorbed by an
$N_\mathrm{H}$=(1.45$\pm{0.10}$)$\times$10$^{24}$ cm$^{-2}$. We found that the scattering
fraction is less than 1\%, and the reflection fraction is   
R=0.7$^{+0.4}_{-0.5}$. 
The slightly steep photon index, which is also not well constrained
(ranging from 1.9 to 2.9),  is not so striking if we take into account the
complexity of the model.

The Suzaku data confirm the presence of a Fe K$\alpha$ line at  6.37$\pm$0.01
keV ($\sigma$$<$50 eV), but with a lower equivalent width
(EW=450$^{+550}_{-50}$ eV) with respect to the value measured with the {\it XMM--Newton} data,
although consistent considering the errors. In order to investigate the presence
of other possible features, we  inspected the residuals around  6.4 keV with
respect to a single power-law model (see Fig. 5); there are clear residuals,
both red and blue--wards the K$\alpha$ line. Some excesses in the data/model
ratio are present around 5.8 keV (observer's frame) and between 6.5 and 7 keV,
while at E$>$ 7 keV there is a clear drop. While the first excess is most
probably due to a bad subtraction of the  emission lines present in the
calibration source (i.e. Mn K$\alpha1$ at 5.899 keV and Mn K$\alpha2$ at 5.888
keV), the drop at energies larger than 7 keV is most likely due to a high level
of the XIS background. Finally, the excess between 6.5 and 7 keV could suggest
the presence of a second He-- or H--like Fe emission line. In particular, we
found that a second narrow ($\sigma$$<$50 eV) line centered at $\sim$6.92 keV
rest--frame (EW$\sim$100 eV) could be actually present (see Fig. 6) even if it
is not statistically required  ($\chi^{2}$/d.o.f.=255.9/188).  This second line
could be associated both to the Fe XXVI Ly$\alpha$ emission line (rest-frame
energy E=6.96 keV) or to the Fe K$\beta$ emission line (rest-frame energy
E=7.058 keV) or to a combination of these two lines. However, we estimated the
flux of the  6.92 keV line and found that it is about 15\% of the flux of the
K$\alpha$ line, as it is expected for the Fe K$\beta$ emission line (see e.g. 
Leahy \& Creighton \cite{lea}). 

As reported in Table 1,  the best--fit values obtained with both models  (one
and two emission lines) are consistent  within the errors. We  refer to these
models as mildly Compton-thick AGN models. In Table 1, the lines at 
$\sim$6.4 and $\sim$6.9 keV are marked as E$_1$ and E$_2$ respectively. The
2--10 keV flux, corrected for the Galactic absorption along the line of sight,
is F(2--10 keV)=2$\times$10$^{-12}$ erg cm$^{-2}$ s$^{-1}$ and,  once corrected
also for the amount of intrinsic absorption, the deabsorbed 2--10 keV luminosity
of the AGN is 7$\times$10$^{43}$ erg s$^{-1}$.

To further test the above  models, we modeled the two components
associated with cold reflection (the 6.4 keV Gaussian emission and the Compton-scattered continuum
from neutral material) using a single and self-consistent model:  {\it reflionx}
(Ross \& Fabian \cite{ross}). In this model, an optically thick disc is
illuminated by a power--law, producing florescence lines and continuum emission.
The parameters include the Fe abundance, ionization parameter $\xi$ (defined as
$\xi$ = 4$\pi$Ftot/n$_H$, where Ftot is the total illuminating flux and n$_H$ is
the density of the reflector) and the incident power-law photon index $\Gamma$. 
Being the observed Fe iron K$\alpha$ emission line unresolved in the Suzaku
spectrum, no additional velocity broadening was applied to the reflected
spectrum. We found that the data are
well reproduced by this model ($\chi^{2}$/d.o.f.=226.3/190), confirming the
physical consistence between the measured reflected continuum and the line
intensity. In this case the reflected component is absorbed by
$N_\mathrm{H}$=(2.7$\pm{0.8}$)$\times$10$^{23}$ cm$^{-2}$, while the transmitted component is
absorbed by  a larger column density $N_\mathrm{H}$=(1.3$\pm{0.3}$)$\times$10$^{24}$
cm$^{-2}$. We found a value for the photon index that is consistent with the
previous ones ($\Gamma$=2.5$\pm$0.5) and a ionization parameter
$\xi$=29$^{+23}_{-9}$ erg cm s$^{-1}$ (the lowest value allowed by the model is
$\xi$=10 erg cm s$^{-1}$), in agreement with iron atoms typically in a
low--ionization state corresponding to Fe I--XVII. Best fit parameters are
reported in Table~1 (mildly Compton-thick AGN models - {\it reflionx}).

Finally, we checked the self-consistence of the mildly Compton--thick hypothesis
by using also the recent paper published by Murphy \& Yaqoob (\cite{murphy}).
They calculated Green's functions that may be used to produce
spectral-fitting routines to model the putative neutral toroidal X-ray
reprocessor in AGNs for an arbitrary input spectrum. In their calculation the
reprocessed continuum and fluorescent line emission due to Fe K$\alpha$, Fe
K$\beta$ and Ni K$\alpha$ are treated self-consistently. On the basis of the EW
obtained for the  Fe K$\alpha$ emission line ($\sim$450 eV) by our analysis and
by assuming an inclination angle between the observer's line of sight  and the
symmetry axis of the torus larger than 60$^{\circ}$, the model predicts an
$N_\mathrm{H}$$\simeq$1.5--3$\times$10$^{24}$ cm$^{-2}$. The mildly Compton--thick
hypothesis is thus fully supported also by this recent
model proposed in the case of neutral toroidal X-ray reprocessor in AGNs. 

{\bf Heavily  Compton-thick hypothesis:} For completeness, we tested   also the
heavily ($N_\mathrm{H}$$>$10$^{25}$ cm$^{-2}$) Compton--thick AGN
hypothesis. In this case, the emission is completely dominated by the scattered
component at low energies and only by the reflected component ({\it reflionx} or
{\it pexrav} model)  at high energies. By  using disc-reflection  models ({\it
reflionx} or {\it pexrav} models), we found that this scenario is statistically
acceptable. As an example, we report in Fig. 7 and Table 1 the results found
with the {\it reflionx} model ($\chi^{2}$/d.o.f. = 218.9/189, 
heavily Compton-thick AGN models - {\it reflionx} in Table 1). However,
an intrinsic column density higher  than 10$^{25}$ cm$^{-2}$ is not  supported
by the model of Murphy \&  Yaqoob (\cite{murphy}) for Fe K$\alpha$ EW of order 
of 450 eV. Moreover, on the basis of this model, the observed luminosity should
be just  less than few percents of the intrinsic one, by implying an intrinsic luminosity of
L(2--10 keV)$\geq$10$^{45}$ erg s$^{-1}$. This value of X--ray 
luminosity exceeds the infrared luminosity measured for this source 
(L$_{IR}$=2.7$\times$10$^{44}$ erg s$^{-1}$), which,  under the assumption that
most of the optical and ultraviolet radiation is absorbed by a dusty torus
surrounding the nuclear source, is  a good proxy of the bolometric luminosity.
Therefore, on the basis of the Murphy \&  Yaqoob (\cite{murphy}) model,
this scenario is not acceptable from the physical point of view.

\begin{figure}[]
\begin{center}
\label{Fig. 5}
\psfig{figure=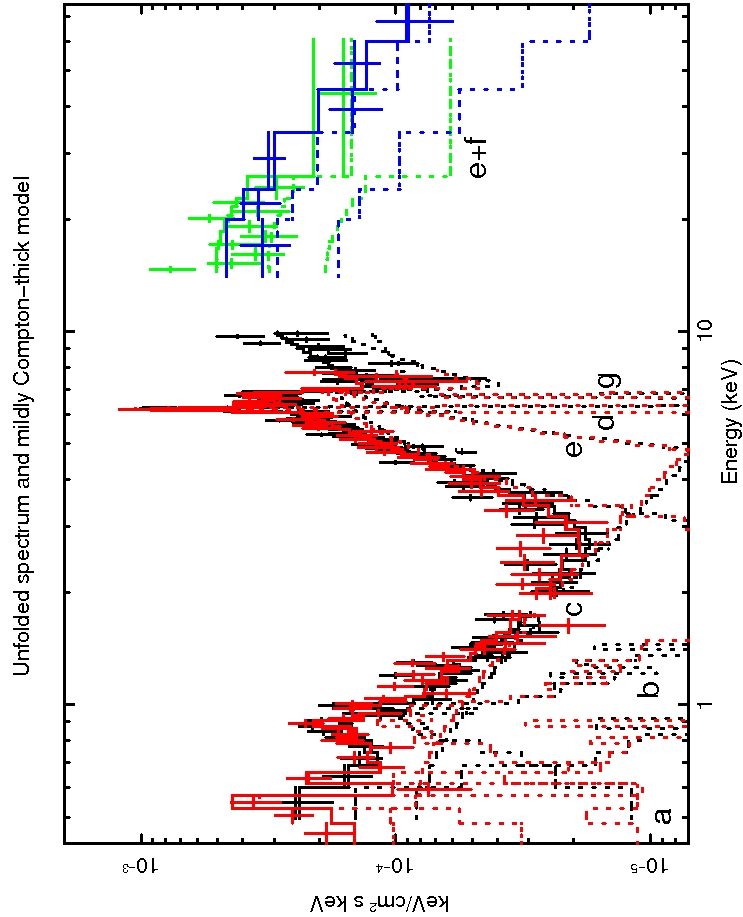,height=9cm,width=7cm,angle=-90}
\psfig{figure=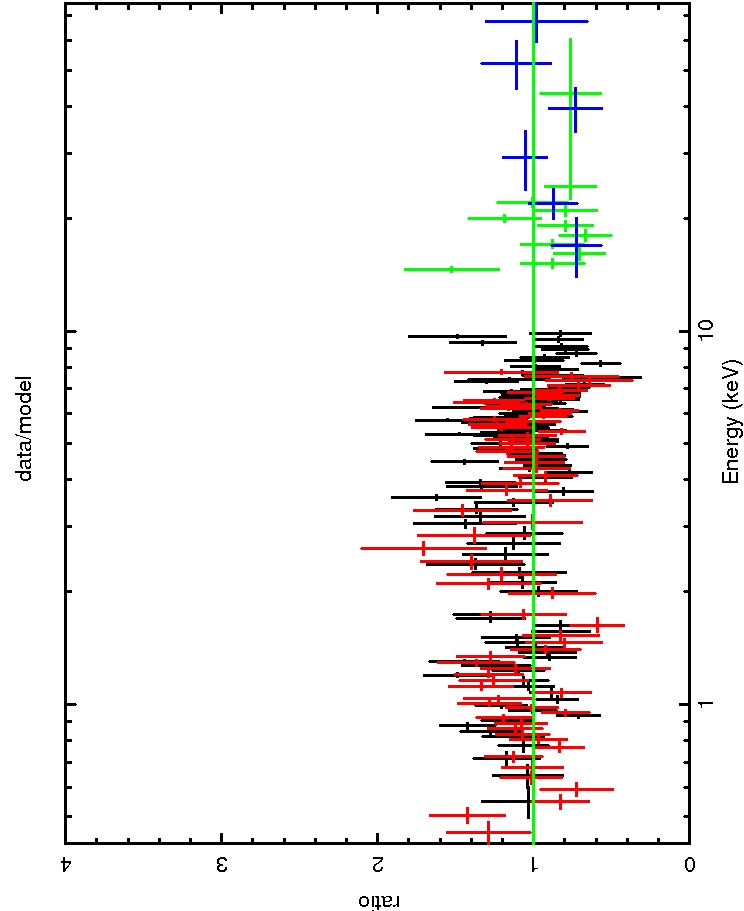,height=9cm,width=6cm,angle=-90}
\caption{Suzaku XIS-FI  (black data points in the electronic version), 
XIS-BI (red in the electronic version), HXD--PIN 
(green in the electronic version) and
SWIFT--BAT  (blue in the electronic version) data of IRAS~04507+0358. {\it Upper panel:} Unfolded
mildly Compton--thick model.
The spectral components are: a) and b) two thermal emission components; c) 
scattered AGN component; d) 
narrow Gaussian line at 6.37 keV; e) absorbed power--law AGN component; 
f) pure reflection AGN component; g) 
narrow Gaussian line at $\sim$6.9 keV.
{\it Lower panel:} Ratio between data and the mildly Compton--thick
best--fit model.} 
\end{center} 
\end{figure}

\begin{figure}[]
\begin{center}
\label{Fig. 6}
\psfig{figure=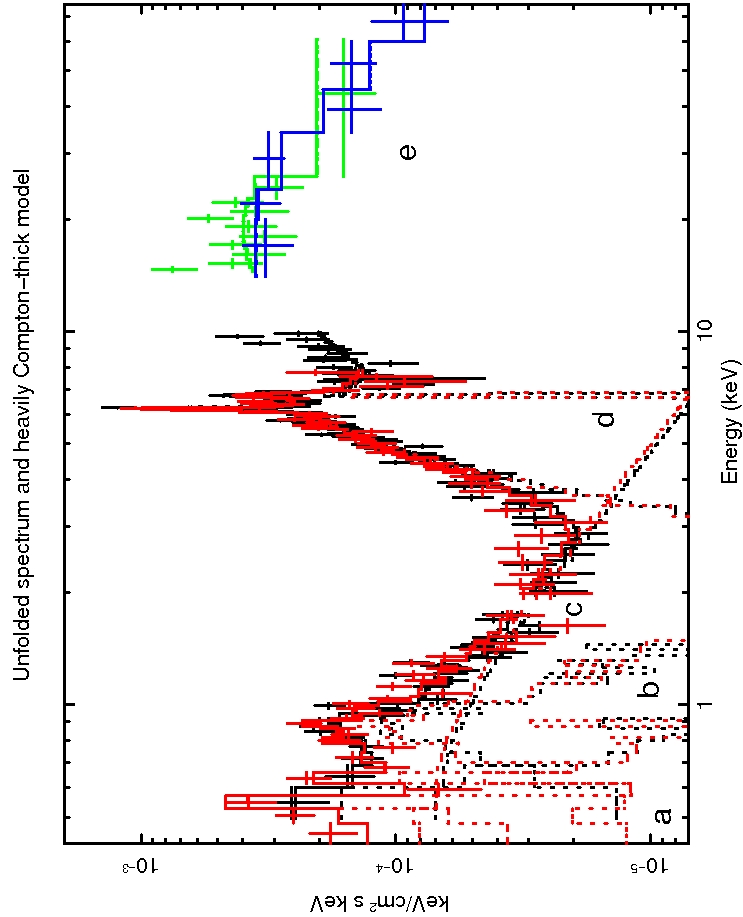,height=9cm,width=7cm,angle=-90}
\psfig{figure=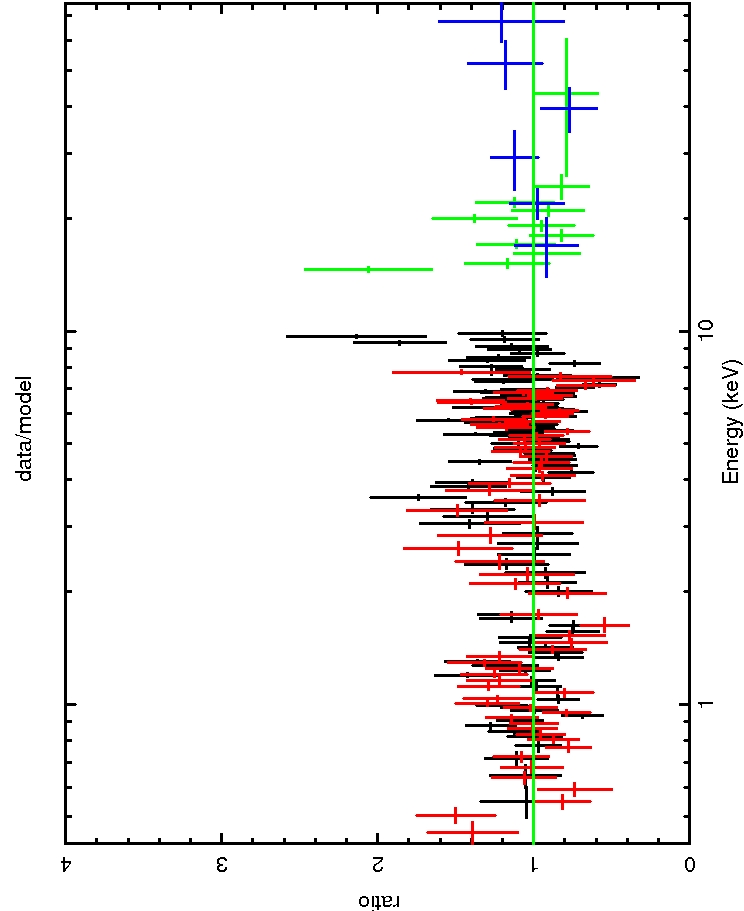,height=9cm,width=6cm,angle=-90}
\caption{Suzaku XIS-FI  (black data points in the electronic version), 
XIS-BI  (red in the electronic version), HXD--PIN (green in the 
electronic version) and
SWIFT--BAT (blue in the electronic version) data of IRAS~04507+0358. 
{\it Upper panel:} The unfolded heavily Compton--thick model 
({\it reflionx}, see Table 1) is overplotted
on the data. 
The spectral components are: a) and b) two thermal emission components; c) 
scattered AGN component; d) 
narrow Gaussian line at $\sim$6.9 keV; e) pure reflection AGN component ({\it
reflionx}).
{\it Lower panel:} Ratio between data and the heavily Compton--thick
best--fit model.} 
\end{center} 
\end{figure}

\begin{table*}
\begin{minipage}[t]{1\textwidth}
\caption{Best--fit values of the mildly and heavily Compton-thick AGN models}   
\label{table:1}        
\renewcommand{\footnoterule}{}  
\begin{tabular}{l c c c c c c c c c c}   
\hline\hline                  
  \multicolumn{11}{c}{\it Mildly Compton-thick AGN models} \\
\hline  
Model & $\Gamma$ & $N_\mathrm{H}$ & Refl. & Scatt. &
E$_1$ & $\sigma$ & E$_2$ & $\sigma$ & 
$\chi^{2}$/d.o.f. & L(2--10 keV) \\ 
&  & [10$^{22}$ cm$^{-2}$]  & frac. & frac.& keV & eV & keV & eV & & 10$^{42}$ erg s$^{-1}$\\  
\hline    
~~\\
One line & 2.4$\pm$0.5 & 145$\pm$10 & 0.7$^{+0.4}_{-0.5}$ & $<$1\% 
& 6.37$\pm{0.01}$ & $<$50  & -- & -- & 265.5/190 & 70 \\   
Two lines & 2.4$\pm$0.5 & 146$^{+7}_{-11}$ & 0.7$^{+0.4}_{-0.5}$ & $<$1\% & 6.37$\pm0.01$ &  $<$50 & 
6.92$^{+0.50}_{-0.20}$ & $<$50 & 255.9/188 & 70 \\  
~~\\
\hline    
Model & $\Gamma$ & $N_\mathrm{H}$ & $\xi$ & E$_2$ & $\sigma$ & $\chi^{2}$/d.o.f. & L(2--10 keV) \\ 
&  & [10$^{22}$ cm$^{-2}$]  & erg cm s$^{-1}$ & keV & eV & & 10$^{42}$ erg s$^{-1}$\\  
\hline    
~~\\ 
Reflionx & 2.5$\pm$0.5 & 130$\pm$30 & 29$^{+23}_{-9}$ & 6.91$^{+0.12}_{-1.71}$ & $<$50 & 226.3/190 & 50\\
~~\\   
\hline\hline
  \multicolumn{11}{c}{\it Heavily Compton-thick AGN model} \\
\hline
Model & $\Gamma$ &  $N_\mathrm{H}$ & $\xi$ & E$_2$ & $\sigma$ & $\chi^{2}$/d.o.f. & L(2--10 keV) \\ 
&   & [10$^{22}$ cm$^{-2}$] & erg cm s$^{-1}$ & keV & eV & & 10$^{42}$ erg s$^{-1}$\\  
\hline
~~\\
Reflionx & 2.2$\pm{0.5}$ & $>$1000\footnote{This value of  $N_\mathrm{H}$ is not an
output of the fit, but it is an assumption that we made in this model. } &
66$^{+10}_{-7}$ &  6.93$^{+0.07}_{-0.08}$ & $<$50 & 218.9/189 & $>$10$^3$ \\ 
~~\\
\hline   
\hline                               
\end{tabular}
\end{minipage}
\end{table*}

\section{Discussion and Conclusion}

IRAS 04507+0358 was proposed as a new Compton--thick AGN candidate through the
diagnostic diagram described in Severgnini et al. (\cite{sev}). Here, by studying
its broad band X--ray spectral properties (0.4-100 keV), we confirm the
Compton--thick nature of this AGN. 

We showed in this paper that with the {\it XMM-Newton} and XIS data below 10 keV
alone, we could not distinguish between Compton-thin or Compton-thick scenarios.
However, the Compton-thin model clearly  under-predicts higher energy data.  In
particular,  we found that the most favorite scenario is that of a mildly
Compton--thick AGN with  $N_\mathrm{H}$=1.3-1.5$\times$10$^{24}$ cm$^{-2}$ and  L(2--10
keV)=5-7$\times$10$^{43}$ erg s$^{-1}$. This luminosity is in full agreement
with the observed infrared luminosity  and with that predicted  for the torus by
Fraquelli et al. (\cite{fraquelli}). These authors, after having estimated the
rate of the ionizing photons emitted by the AGN and assuming an opening angle of
30$^\circ$ for the ionization cones, predicted an infrared luminosity for the
torus of L$_{IR}$=5.3$\times$10$^{43}$ erg s$^{-1}$.  This value is about
20\% of the  infrared luminosity (L$_{IR}$$\simeq$2.7$\times$10$^{44}$
erg s$^{-1}$, see  also Sect. 2). The remaining fraction of the infrared
luminosity (2.2$\times$10$^{44}$ erg s$^{-1}$) is most probably due to
star-formation activity and it is roughly in agreement with the soft X--ray
luminosity of the thermal components of our source (L(0.5--2
keV)=2.5$\times$10$^{41}$ erg s$^{-1}$) considering the L(soft--X)-L(FIR) 
relation presented in Persic et al. (\cite{persic}). 
To estimate the relevant SFR of our source, we adopted the following
Kennicutt et al. (\cite{kennicutt98}) relation: SFR=(L$_{FIR}$/5.8$\times$10$^9$
L$_\odot$) M$_\odot$ yr$^{-1}$ and we considered only the infrared luminosity
produced by star--formation activity ($\sim$2.2$\times$10$^{44}$ erg s$^{-1}$).
We found a SFR of about 10 M$\odot$/yr.

As for the heavily Compton--thick hypothesis,  by using disc--reflection models
it is possible to well reproduce the broad--band spectrum of IRAS 04507+0358 and
to obtain a statistically acceptable fit of the data. 
In this case, in order not to exceed the bolometric luminosity of the source,
the reflection component should be less than 10\% of the
intrinsic one, which is not supported by the 
Murphy \&  Yaqoob (\cite{murphy}) model.

We note that this source could not be found as a possible
Compton-thick AGN by using only the data below 10 keV, nor by using the 
two-dimensional diagnostic tool  for  reflection-dominated Compton-thick
object proposed by Bassani et al. (\cite{bass99}). This latter diagram shows
that Compton-thick AGN should be characterized by  high K$\alpha$ iron line
equivalent width (EW$>$300 eV) and by a 2-10 keV flux normalized to the [OIII]
optical-line flux  (T parameter)  typically lower than 0.5. In this estimate,
the [OIII] optical-line flux should be corrected for Galactic and intrinsic
extinction. From the X--ray analysis performed on IRAS 04507+0358, we derived for
the Fe K$\alpha$ line an EW of $\sim$450 eV. In order to estimate the T parameter, we
considered  the [OIII] line flux  (already corrected for the Galactic
extinction) reported by Cid Fernandes et al. (\cite{cid}). 
After the correction for the intrinsic extinction\footnote{For the intrinsic extinction, we
applied the following formula: Fcor[OIII] =
Fobs[OIII][(H$\alpha$/H$\beta$)obs/(H$\alpha$/H$\beta$)$_0$]$^{2.94}$ (Bassani
et al. \cite{bass99}), where Fcor[OIII] is the extinction-corrected flux of
[OIII]5007, and Fobs[OIII] the observed flux of [OIII]5007. We used and observed
line ratio of  H$\alpha$/H$\beta$=5.5  (Cid Fernandes et al. 2001) and  an
intrinsic Balmer decrement of (H$\alpha$/H$\beta$)$_0$=3.0.}, we found T=2,
larger than the typical value adopted to select Compton-thick candidates
($<<$1). This implies that by using 2--10 keV data alone and/or diagnostic
diagrams
based on the Fe K$\alpha$ line EW vs. 2--10 keV to [OIII] flux ratio, a
significant fraction of mildly or heavily Compton--thick AGN could be missed.
Whereas some examples of Compton--thick AGN with low  value of Fe K$\alpha$ line
equivalent width are indeed already present in the literature (e.g. Awaki et al.
\cite{awaki2000}; Ueda et al. \cite{ueda}; Braito et al. \cite{braito09}), IRAS
04507+0358 escapes the diagnostic diagram due to its high value of T. In the case of
mildly Compton--thick AGN, this is most
probably due to the large amount of light reflected from the torus (i.e.
about 35\% of the light emitted by  the central source) that increases the
continuum in the 2-10 keV range.  The Compton--thick AGN missed using the
Bassani et al. (\cite{bass99}) criteria could be selected by combining infrared
information with 2-10 keV data and with very hard X--ray follow--ups, as shown
by the study presented here. Moreover, our analysis, besides providing a further
evidence of the potentiality of the F(2-10 keV)/F24$\mu$m vs. HR diagnostic
diagram, also shows the importance  of using a wide X--ray spectral coverage 
in order to constrain the intrinsic column density in this type of sources even in the
case of mildly Compton-thick AGN. 

With the aim of establishing the nature of other Compton-tick
AGN candidates found through our diagram, we have recently obtained in the
Suzaku AO5 call 100 ksec of observation for MCG-03-58-007, while the broad--band
X--ray analysis for other twelve sources with a SWIFT--BAT counterpart is
already ongoing and will be the subject of a forthcoming paper. 
The final aim is to better constrain and define the
Compton--thick AGN population and hence to better estimate their space density.

\begin{acknowledgements}
The authors acknowledge financial support from ASI 
(grant n. I/088/06/0, COFIS contract and grant n. I/009/10/0).
VB acknowledge support from the UK STFC research council.
We thank Giancarlo Cusumano for his useful support with the SWIFT--BAT data. 
We also thank the anonymous referee for her/his useful
comments.

\end{acknowledgements}

\end{document}